\documentclass[prl,aps,twocolumn,amsfonts]{revtex4}

\usepackage{graphicx}
\usepackage{epsfig}
\def\qed{\leavevmode\unskip\penalty9999 \hbox{}\nobreak\hfill
     \quad\hbox{\leavevmode  \hbox to.77778em{%
               \hfil\vrule   \vbox to.675em%
               {\hrule width.6em\vfil\hrule}\vrule\hfil}}
     \par\vskip3pt}

\def\ibb #1{\leavevmode\hbox{\kern.3em\vrule
     height 1.5ex depth -.1ex width .4pt\kern-.3em\rm#1}}

\newcommand{\be}[1]{\begin{equation} #1 \end{equation}}
\newcommand{\bea}[1]{\begin{eqnarray} #1 \end{eqnarray} }
\newcommand{\ba}[2]{\left(\begin{array}{#1}#2\end{array}\right)}
\newcommand{\tr}[1]{{\rm Tr}\left(#1\right)}
\newtheorem{theorem}{Theorem}
\newtheorem{lemma}{Lemma}

\begin{document}

\title{Entanglement versus Bell violations\\and their behaviour under local filtering operations}

\author{Frank Verstraete}\email{frank.verstraete@rug.ac.be}
 \affiliation{Department of Mathematical Physics and Astronomy, Ghent University, Belgium\\
 Department of Electrical Engineering (SISTA), KULeuven,  Belgium}
\author{Michael M. Wolf}
 \email{mm.wolf@tu-bs.de}
 \affiliation{Institute for Mathematical Physics, TU Braunschweig, Germany}

\date{\today}

\begin{abstract}
We discuss the relations between the violation of the CHSH Bell
inequality for systems of two qubits on the one side and
entanglement of formation, local filtering operations, and the
entropy and purity on the other. We calculate the extremal Bell
violations for a given amount of entanglement of formation and
characterize the respective states, which turn out to have
extremal properties also with respect to the entropy, purity and
several entanglement monotones. The optimal local filtering
operations leading to the maximal Bell violation for a given state
are provided and the special role of the resulting Bell diagonal
states in the context of Bell inequalities is discussed.
\end{abstract}

 \maketitle

\section{Introduction and preliminaries}
Entanglement has always been a key issue in the ongoing debate
about the foundations and interpretation of quantum mechanics
since Einstein, Podolsky and Rosen (EPR) published their famous
gedanken-experiment in 1935 \cite{EPR}. For a long time
discussions about entanglement were purely meta-theoretical.
However, this appeal was changed dramatically in 1964 by John
Bell's \cite{Bell64} observation that the EPR dilemma could be
formulated in the form of assumptions naturally leading to a
falsifiable prediction. The experimental fact that these {\it Bell
inequalities} can indeed be violated \cite{Aspect} has not only
ruled out a single theory, but the very way theories had been
formulated for quite a long time. Whereas until 1989 entanglement
was widely believed to be equivalent to the violation of a Bell
inequality, it turned out that such a violation is neither
necessary for mixed entangled states \cite{Werner89} nor a good
measure for the amount of entanglement \cite{Popescu, Gisin}. It
was in particular shown by Gisin \cite{Gisin} that some states
initially satisfying Bell's inequalities, lead to a violation
after certain local filtering operations. Hence, local filtering
operations can on an average increase the degree of violation
while decreasing the amount of entanglement. Although general
structural knowledge about entanglement \cite{entreview} on the
one side and Bell's inequalities \cite{Bellreview} on the other
has increased dramatically in the last few years, our knowledge
about their relation is still mainly restricted to the fact that
states violating a Bell inequality have to be entangled.

The present paper is devoted to settling the relationship between
entanglement, measured in terms of the concurrence, the Bell
violations and their behaviour under local filtering operations
for the case of two qubit systems.

To fix ideas we will start by recalling some of the basic
definitions and properties. Throughout this paper we will consider
systems of two qubits -- one may think of two spin $\frac12$
particles or the polarization degrees of freedom of two photons --
for which we can explicitly calculate the amount of entanglement,
in terms of the {\it entanglement of formation} \cite{Wootters},
as well as the maximal violation of the Bell inequality in its
Clauser-Horne-Shimony-Holt (CHSH) form \cite{CHSH,HHH}.

The concept of entanglement of formation ($EoF$) is related to the
amount of entanglement needed to prepare the state $\rho$, and it
was shown by Wootters \cite{Wootters} that
\begin{equation}\label{EoF}
EoF(\rho) = h\left(\frac{1+\sqrt{1-C^2}}2\right),
\end{equation}
where $h(x)=-x\lg{x}-(1-x)\lg{(1-x)}$ and the {\it concurrence}
$C=\max\left[0, \sqrt{l_1}-\sum_{i=2}^4\sqrt{l_i}\right]$ with
$\{l_i\}$ being the decreasingly ordered eigenvalues of
$\rho(\sigma_y\otimes\sigma_y)\rho^T(\sigma_y\otimes\sigma_y)$. In
order to circumvent lengthy logarithmic expressions we will in
following use the concurrence rather than $EoF$, which is in fact
a convex and monotone function with respect to $C$.

The CHSH inequality formulated for two qubit systems states, that
within any local classical model the expectation value
$\tr{\rho{\cal B}}$ of the {\it Bell operator}
\begin{equation}\label{BI}{\cal
B}=\frac12 \sum_{ij=1}^3\big[a_i (
c_j+d_j)+b_i(c_j-d_j)\big]\sigma_i\otimes\sigma_j,
\end{equation}
where $(\vec{a},\vec{b},\vec{c},\vec{d})$ are real unit vectors
and $\sigma_i$ being the Pauli matrices, has to be bounded by one.
Its violation is a measure of how strong non-classical properties
of the state manifest themselves in correlation experiments.

In the sequel we will often represent the two qubit state in terms
of the $4\times 4$ matrix
$\tilde{R}_{ij}=\tr{\rho\sigma_i\otimes\sigma_j}$ (with $\sigma_0$
being the identity) and the $3\times 3$ block
$R_{kl}=\tr{\rho\sigma_k\otimes\sigma_l}$, where $k,l=1,2,3$. It
is important to note, that the latter can be diagonalized just by
changing the local bases, which will neither affect the
entanglement nor the maximal Bell violation.

\section{Entanglement of formation versus violation of the CHSH inequality}

In \cite{HHH}the Horodecki family showed, that the maximal
violation of the CHSH inequality can be calculated by considering
the $3\times 3$ matrix $R_{kl}=\tr{\rho\sigma_k\otimes \sigma_l}$.
We will give an alternative derivation of this result in a way
that will be very useful in the sequel:
\begin{lemma}\label{lemmaHor}(Horodecki \cite{HHH}) Given the decreasingly ordered
singular values $\{\sigma_i\}$ of $R$, the maximal violation
$\beta(\rho)=\max_{\cal B} \tr{\rho{\cal B}}$ is given by
$\sqrt{\sigma_1^2+\sigma_2^2}$.
\end{lemma}
{\em Proof:} Translated into the $R$-picture, calculating the
maximal expectation value of $\cal B$ under the constraint that
$(\vec{a},\vec{b},\vec{c},\vec{d})$ are real unit vectors, amounts
to maximizing  $\tr{R X}$  with
\be{X=\ba{cc}{\vec{c}&\vec{d}}\frac{1}{2}\ba{cc}{1&1\\1&-1}\ba{c}{\vec{a}^T\\
\vec{b}^T}.} It is an easy exercise to show that $X$ is a real
$3\times 3$ matrix, subjected to the only constraints that it be
of rank $2$ and that $\tr{X^T X}=1$. Standard linear algebra then
dictates that $\tr{RX}$ is maximized iff $X$ is chosen to be
proportional to the best rank 2 least-squares approximation of the
matrix $R$. In the basis where $R$ is diagonal ($R={\rm
diag}(\sigma_1,\sigma_2,\sigma_3)$), $X$ is therefore given by
$X={\rm diag}(\sigma_1,\sigma_2,0)/\sqrt{\sigma_1^2+\sigma_2^2}$,
which immediately leads to
$\beta=\sqrt{\sigma_1^2+\sigma_2^2}$.\qed

\begin{figure}\epsfig{file=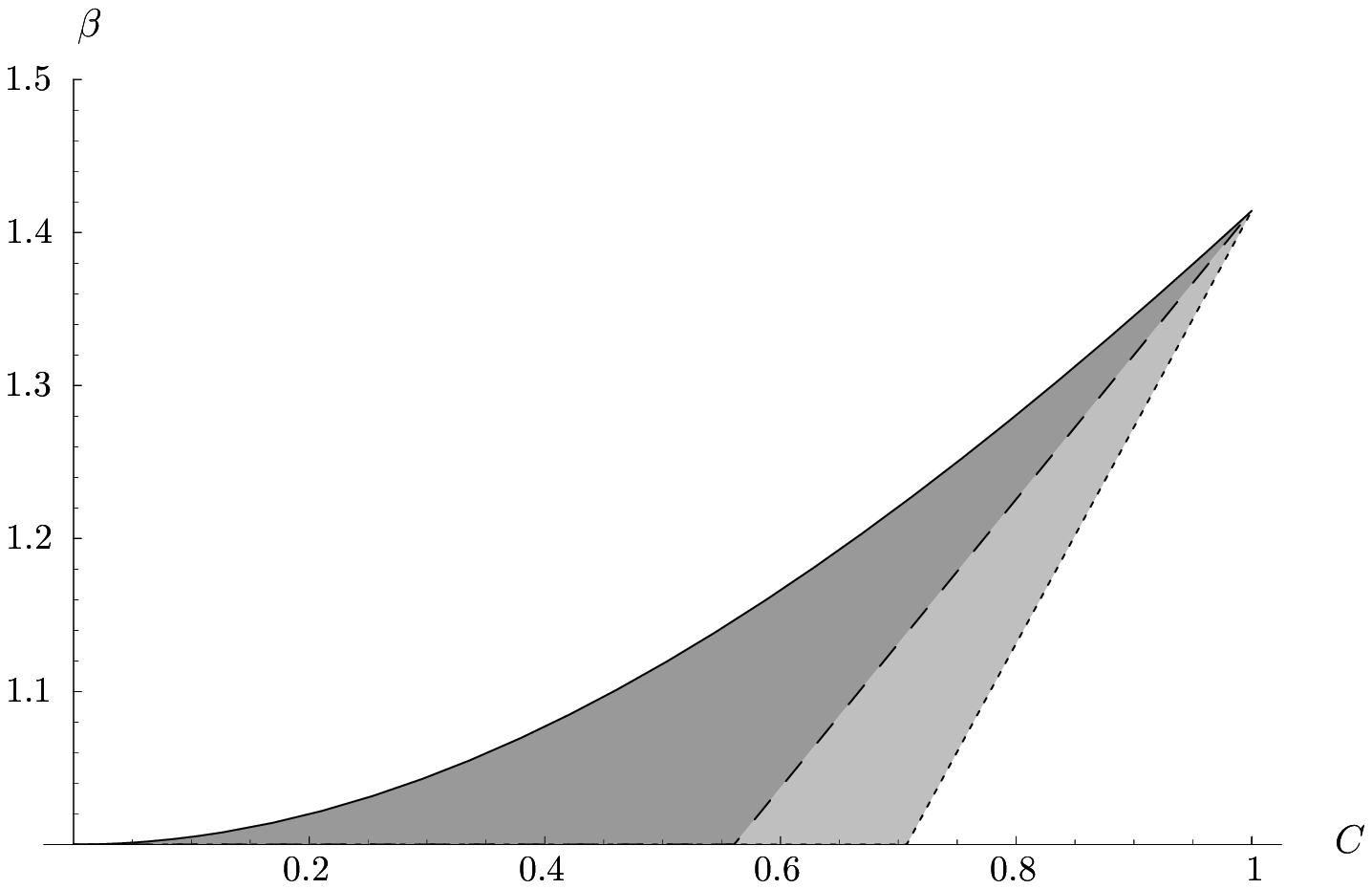,width=8.5cm}
\caption{\label{fig:epsart} The region of possible maximal Bell
violation for given concurrence. The dark grey region corresponds
to Bell diagonal states and the three lines represent pure states
(solid), Werner states (dashed) and maximally entangled mixed
states (dotted).}
\end{figure}

In the following we will derive the extremal violations for a
given amount of entanglement plotted in Fig.\ref{fig:epsart}.

\begin{theorem}\label{theomax}
The maximal violation of the CHSH inequality for given concurrence
$C$ is $\beta(\rho)=\sqrt{1+C^2(\rho)}$.
\end{theorem}
{\em Proof:} As shown by Wootters \cite{Wootters}, it is possible
to decompose a mixed state of two qubits $\rho = \sum_i p_i
|\psi_i\rangle\langle \psi_i|$ into a convex sum of pure states,
all with concurrence equal to the concurrence of the mixed state.
Since the extremal violation is moreover a convex function, i.e.,
$\max_{\cal B}\tr{\rho{\cal B}}\leq \sum_i p_i \max_{\cal
B}\langle\psi_i |{\cal B}|\psi_i\rangle$, it is sufficient to have
a look at pure states, which can always be written in their
Schmidt form as $|\psi\rangle=\lambda_+
|00\rangle+\lambda_-|11\rangle$ with
$\lambda_{\pm}=(\sqrt{1+C}\pm\sqrt{1-C})/2$. The corresponding
$R$-matrix is diagonal with singular values $(1,C,C)$ leading to
$\beta=\sqrt{1+C^2}$.\qed

It is interesting to note that there also exist mixed states of
rank 2 for which the violation is as strong as for pure states.
These are, up to local unitary operations, all of the form
\be{\rho=\frac{1}{2}\ba{cccc}{.&.&.&.\\.&1-a&C&.\\.&C&1+a&.\\.&.&.&.}}
with $C$ being the concurrence and $a$ a free real parameter
constrained by $|a|\leq\sqrt{1-C^2}$, where equality leads to pure
states and {\it Bell diagonal states} (see section below) are
obtained for $a=0$.

\begin{theorem}\label{theomin}
The minimal violation of the CHSH inequality for given concurrence
$C$ is given by $\beta(\rho)=\max[1,\sqrt{2}C(\rho)]$.
\end{theorem}
The proof is quite technical and may be skipped by readers not
interested in technical details.\\
{\em Proof:} We will use similar techniques as used in
\cite{VDD01a,VADD01,VDD01d}, where it was shown that surfaces of
constant concurrence can be generated by transforming
$\tilde{R}\mapsto \tilde{R}'=L_1 \tilde{R}L_2^T$ by left and right
multiplication with proper orthochronous Lorentz transformations,
taken into account the constraint that the $(0,0)$ element of
$\tilde{R}$ (representing the trace of $\rho$) does not change
under these transformations. They leave the Lorentz singular
values \cite{VDD01a} invariant, and the concurrence is a function
of these four parameters only.

Using the variational characterization used in lemma
\ref{lemmaHor}, the first step consists of varying the Lorentz
transformations $L_1,L_2$ and the $3\times 3$ rank 2 matrix  $X$
(with constraint $\tr{X^T X}=1$), and imposing that these
variations be zero (i.e. we have an extremum). The object function
is given by
\be{\tr{L_1\tilde{R}L_2^T\ba{cc}{0&0\\0&X}}\label{kost1}} under
the constraints $\tr{X^T X}=1$ and \be{
\tr{L_1\tilde{R}L_2^T\ba{cccc}{1&.&.&.\\.&.&.&.\\.&.&.&.\\.&.&.&.}}=1.}
The orthogonal degrees of freedom of $X$ can be absorbed into
$L_1,L_2$, such as to yield a diagonal $X$ of rank 2: $X={\rm
diag}(q,r,0)$ with $q^2+r^2=1$. Variation of the Lorentz
transformations yields the extremal
conditions\be{\tr{G_1\tilde{R}'\ba{cc}{\lambda&0\\0&X}}
=\tr{\tilde{R}'G_2\ba{cc}{\lambda&0\\0&X}}=0.} for all possible
generators $G_1,G_2$ of the Lorentz group and $\lambda$ being a
Lagrange parameter. The generators are all of the form
\be{\label{gen}G=\ba{cc}{0&\vec{v}\\ \vec{v}^T & A}} with
$\vec{v}\in\mathbb{R}^3$ and $A$ a real and antisymmetric $3\times
3$ block. A detailed discussion of the case $\lambda\neq 0$ shows,
that this leads to the maximal violation, which we have already
obtained in theorem \ref{theomax} using a more simple
argumentation. The minimal value of the Bell violation turns out
to correspond to the case where $\lambda=0$ and yields the
condition that $\tilde{R}'$ is of form
\be{\tilde{R}'=\ba{cccc}{1&.&.&a\\.&x&.&.\\.&.&y&.\\
b&.&.&z}.} The extremal violation of the Bell inequality is then
directly found by varying the remaining diagonal elements of $X$,
leading to a violation given by $\sqrt{x^2+y^2}$. The concurrence
of the extremal state can be calculated explicitly, and is given
by:
\bea{C&=&\frac12 \max\big[0,|x-y|-\sqrt{(1-z)^2-(a-b)^2},\nonumber\\
&&\hspace{.6cm}|x+y|-\sqrt{(1+z)^2-(a+b)^2}\big].\label{minR}} The
constraints that $\tilde{R}$ corresponds to a (positive) state
are expressed by the inequalities \bea{-1\leq &z&\leq 1 ,\\
(1+z)^2-(a+b)^2&\geq&(x-y)^2 ,\\
(1-z)^2-(a-b)^2&\geq&(x+y)^2.} Applying these to the expression of
the concurrence, this immediately leads to the sharp inequality
$C\leq\min(|x|,|y|)$. The Bell violation, given by
$\beta=\sqrt{x^2+y^2}$, will then be minimal for given concurrence
if $|x|=|y|$, leading to final result:
$\beta(\rho)\geq\sqrt{2}C(\rho)$. To complete the proof, we still
have to check if there indeed exists a state with the properties
that $x=y$, $(1+z)^2=(\alpha+\beta)^2$,
$(1-z)^2-(\alpha-\beta)^2\geq(x+y)^2$,$-1\leq z\leq 1$ and
$|z|\leq |x|$. Choosing for example $\alpha=\beta=(1+z)/2$ and
$z=-|x|$ indeed leads to a possible result, which is a convex
combination of a maximally entangled and an orthogonal separable
pure state. Note that all parameters fulfilling the above
constraints lead to states with the same minimal possible amount
of $\beta$ for given concurrence.\qed

The states minimizing the Bell violation for given entanglement of
formation are all rank deficient and belong to the class of
maximally entangled mixed states introduced by Ishizaka and
Verstraete et al. \cite{IH00,VAD01}. These states do have a
remarkable property: their entanglement of formation, negativity
\cite{VW01} and relative entropy of entanglement \cite{VP98}
cannot be increased by any global unitary transformation
\cite{VAD01} (and thus under any transformation preserving the
spectrum). For given entanglement their entropy is the largest
 and their purity (measured in terms of $\tr{\rho^2}$) is the
 smallest possible one. In \cite{VADD01} it was shown that these states also minimize
the negativity and the relative entropy of entanglement for a
fixed amount of entanglement of formation, which is fully
compatible with the result about the minimal violation of the Bell
inequality.

\section{Optimal Filtering}
Local filtering operations on single copies are of particular
importance whenever it is difficult or even impossible to operate
jointly on several copies -- such as in single photon experiments.
Gisin \cite{Gisin} noted that there exist mixed states that do not
violate any CHSH inequality but can violate them after a filtering
operation is applied to them. Therefore the question is raised:
what local filtering operation has to be applied to a given state
such as to yield a new state that violates the CHSH inequality
maximally?

\begin{theorem}\label{theofilter}
Given a single copy of a state $\rho$, then the optimal local
filtering operations yielding a state with maximal possible
violation of the CHSH inequality are the unique stochastically
reversible filtering operations bringing the state into Bell
diagonal form.
\end{theorem}

{\em Proof:} The proof is completely similar to the proof of
theorem \ref{theomin}, so we will only repeat the major steps. In
the $\tilde{R}$-picture, filtering operations correspond to left
and right multiplication with Lorentz transformations, followed by
renormalization. The function, which we have to maximize with
respect to $L_1, L_2$ and $X={\rm diag}(q,r,0)$ in order to obtain
the maximal Bell violation, therefore becomes\be{\label{beta3}
\tr{\frac{L_1\tilde{R}L_2^T}{\left(L_1\tilde{R}L_2^T\right)_{00}}\ba{cc}{0&0\\0&X}}}
with the constraint $q^2+r^2=1$ and the normalization factor
$(L_1\tilde{R}L_2^T)_{00}$. Variation leads to the condition
\[\tr{G_1\tilde{R}'\ba{cc}{-\beta&0\\0&X}}
=\tr{\tilde{R}'G_2\ba{cc}{-\beta&0\\0&X}}=0,\] where again this
has to hold for arbitrary $G_1,G_2$, and where $\beta$ is equal to
Eq. (\ref{beta3}), i.e., the Bell expectation value for given
$q,r,L_1,L_2$. If $\beta>1$ (i.e. Bell violation), it holds that
$\beta$ cannot be equal to $|q|$ or $|r|$, and the form of the
generators in Eq.(\ref{gen}) implies that the above equations can
only be satisfied iff $\tilde{R}'$ is diagonal corresponding to a
Bell diagonal state (see next section). In \cite{VDD01a} it was
shown that for each mixed state there exist local filtering
operations bringing the state into a unique Bell diagonal form,
such that we have proven that these are the filtering operations
that maximize the Bell violation. For a more detailed discussion
of these filtering operations we refer to Ref. \cite{VDD01a}. \qed

This result was expected as it was shown in \cite{VDD01a} that
exactly the same filtering operations maximize the entanglement of
formation and the negativity.

Theorem \ref{theofilter} implies that there exists a large number
of mixed entangled states that do not violate any CHSH inequality.
A specific example was already given by Werner \cite{Werner89},
and here we have shown that whatever state whose Bell diagonal
normal form does not violate the CHSH inequalities cannot violate
any CHSH inequality, even after all possible local filtering
operations.

In the following section we will discuss in more detail the {\it
Bell diagonal states}, for which theorem \ref{theofilter} shows
that the Bell violation cannot be increased by any local filtering
operation.

\section{The role of Bell diagonal states}

We call a state {\it Bell diagonal} if there is a local choice of
bases such that it can be written as a convex combination of the
four maximally entangled Bell states \cite{BVSW}, which means that
$\tilde{R}$ is diagonal in that basis. The diagonal elements of
$R$ then only depend on the eigenvalues
$\lambda_1\geq\ldots\geq\lambda_4$ of the Bell diagonal state
\cite{VDD01a} and it is thus straight forward to show that the
maximal Bell violation is
\be{\label{Bellbeta}\beta=\sqrt{2}\sqrt{(\lambda_2-\lambda_3)^2+(\lambda_1-\lambda_4)^2}.}
Since the concurrence is given by $C=\max [0,2\lambda_1-1]$ the
region of possible violations is in this case
\be{\sqrt{2}(2C+1)/3\leq\beta\leq\sqrt{1+C^2},} where the lower
bound is sharp for {\it Werner states} \cite{Werner89} and the
upper bound is attained for rank 2 Bell diagonal states and is
equal to the relation for pure states.

The fact that the Bell operator ${\cal B}$ in Eq.(\ref{BI}) is
itself Bell diagonal due to $\tr{{\cal
B}\sigma_i\otimes\sigma_0}=\tr{{\cal B}\sigma_0\otimes\sigma_i}=0$
already suggests that Bell diagonal states play a special role in
the context of violations of the CHSH inequality. And in fact, in
addition to being the optimal outcomes of local filtering
operations, they exhibit another special property:

\begin{theorem}
\label{theoBell}For any given spectrum of the density matrix the
respective Bell diagonal state $\rho$ maximizes the Bell
violation, i.e. $\forall U\in U(4): \beta(\rho)\geq \beta(U\rho
U^*) $.
\end{theorem}
{\em Proof:} First note that as we have to calculate a supremum
over all unitary rotations of the state $\rho$, we can without
loss of generality assume that the initial state commutes with the
Bell operator ${\cal B}$. The proof of the theorem is then based
on the fact that if $u_{ik}$ are the matrix elements of a unitary
matrix, then $|u_{ik}|^2$ is a doubly stochastic matrix, i.e., a
convex combination of permutations $\tau$ (cf.\cite{Bhatia}). If
$\{\lambda_i\}, \{b_i\}$ are the decreasingly ordered eigenvalues
of $\rho$ resp. ${\cal B}$, then
\bea{\tr{U\rho U^*{\cal B}}&=&\sum_{ik}\lambda_i b_k|u_{ik}|^2\ =\ \sum_\tau p_\tau \sum_i \lambda_i b_{\tau (i)}\nonumber\\
&\leq&\sum_i \lambda_i b_i\ =\ \tr{\rho{\cal B}}.}

This immediately implies that if we fix any spectral property of
the state, such as the purity $\tr{\rho^2}$ or the entropy
$-\tr{\rho\log\rho}$, the maximal violation of the CHSH inequality
will always be attained for Bell diagonal states.

\section{Conclusion}
We derived the range of Bell violations for a given amount of
entanglement (measured in terms of the concurrence) and discussed
the extremal states, which turned out to have extremal properties
also with respect to several entanglement monotones, the entropy
and purity. It was conjectured by Munro et al. \cite{Munro} that
for given concurrence the Bell violation increases with the purity
of the state. Although this is not true in general (which can
already be seen from Eq.(\ref{Bellbeta})), our results show that
this is indeed true for the extremal cases.

Moreover, we proved that the local single copy filtering
operations which maximize the concurrence and other entanglement
monotones also maximize the Bell violation and lead to Bell
diagonal states, which in turn are optimal with respect to global
unitary operations as well.

\acknowledgments FV acknowledges the hospitality of the TU
Braunschweig, and is grateful to Jeroen Dehaene for valuable
comments.
\bibliographystyle{unsrt}

\end{document}